# Highly Accurate Log Skew Normal Approximation to the Sum of Correlated Lognormals

Marwane Ben Hcine[1] and Ridha Bouallegue[2]

[1,2]Innovation of Communicant and Cooperative Mobiles Laboratory, INNOV'COM
Sup'Com, Higher School of Communication
Univesity of Carthage
Ariana, Tunisia
[1]marwen.benhcine@supcom.tn
[2]ridha.bouallegue@supcom.rnu.tn

**ABSTRACT**

*Several methods have been proposed to approximate the sum of correlated lognormal RVs. However the accuracy of each method relies highly on the region of the resulting distribution being examined, and the individual lognormal parameters, i.e., mean and variance. There is no such method which can provide the needed accuracy for all cases. This paper propose a universal yet very simple approximation method for the sum of correlated lognormals based on log skew normal approximation. The main contribution on this work is to propose an analytical method for log skew normal parameters estimation. The proposed method provides highly accurate approximation to the sum of correlated lognormal distributions over the whole range of dB spreads for any correlation coefficient. Simulation results show that our method outperforms all previously proposed methods and provides an accuracy within 0.01 dB for all cases.*

## 1. INTRODUCTION

Multipath with lognormal statistics is important in many areas of communication systems. With the emergence of new technologies (3G, *LTE*, *WiMAX*, Cognitive Radio), accurate interference computation becomes more and more crucial for outage probabilities prediction, interference mitigation techniques evaluation and frequency reuse scheme selection. For a given practical case, Signal-to-Interference-plus-Noise (SINR) Ratio prediction relies on the approximation of the sum of correlated lognormal RVs. Looking in the literature; several methods have been proposed in order to approximate the sum of correlated lognormal RVs. Since numerical methods require a time-consuming numerical integration, which is not adequate for practical cases, we consider only analytical approximation methods. Ref [1] gives an extension of the widely used iterative method known as Schwartz and Yeh (SY) method [2]. Some others resources uses an extended version of Fenton and Wilkinson methods [3-4]. These methods are based on the fact that the sum of dependent lognormal distribution can be approximated by another lognormal distribution. The non-validity of this assumption at distribution tails, as we will show later, is the main raison for its fail to provide a consistent approximation to the sum of correlated lognormal distributions over the whole range of dB spreads. Furthermore, the accuracy of each method depends highly on the region of the resulting distribution being examined. For example, Schwartz and Yeh (SY) based methods provide acceptable accuracy in low-precision region of the Cumulative Distribution Function (*CDF*) (i.e., 0.01–0.99) and the Fenton–Wilkinson (FW) method offers high accuracy in the high-value region of the CDF (i.e.,

0.9–0.9999). Both methods break down for high values of standard deviations. Ref [5] propose an alternative method based on Log Shifted Gamma (LSG) approximation to the sum of dependent lognormal RVs. LSG parameters estimation is based on moments computation using Schwartz and Yeh method. Although, LSG exhibits an acceptable accuracy, it does not provide good accuracy at the lower region.

In this paper, we propose a very highly accurate yet simple method to approximate the sum of correlated lognormal RVs based on Log Skew Normal distribution (LSN). LSN approximation has been proposed in [6] as a highly accurate approximation method for the sum of independent lognormal distributions, Furthermore a modified LSN approximation method is proposed in [7]. However, LSN parameters estimation relies on a time-consuming Monte Carlo simulation. The main contribution on this work is to provide a simple analytical method for LSN parameters estimation without the need for a time-consuming Monte Carlo simulation or curve fitting approximation. Our analytical fitting method is based on moments and tails slope matching for both distributions. This work can be seen as extension to the correlated case for our work done in [8].

The rest of the paper is organized as follows: In section 2, a brief description of the lognormal distribution and sum of correlated lognormal distributions is given. In section 3, we introduce the Log Skew Normal distribution and its parameters. In section 4, we use moments and tails slope matching method to estimate LSN distribution parameters. In section 5, we provide comparisons with well-known approximation methods (i.e. Schwartz and Yeh, Fenton–Wilkinson, LSG) based on simulation results.

The conclusion remarks are given in Section 6.

## 2. SUM OF CORRELATED LOGNORMAL RVs

Given $X$, a Gaussian RV with mean $\mu_X$ and variance $\sigma_X^2$, then $L = e^X$ is a lognormal RV with (PDF):

$$f_L(l, \mu_X, \sigma_X) = \begin{cases} \frac{1}{\sqrt{2\pi} l \sigma_X} \exp(-\frac{1}{2\sigma_X^2}[\ln(l) - \mu_X]^2) & l > 0 \\ 0 & \text{otherwise} \end{cases} \quad (1)$$

$$= \phi(\frac{\ln(l) - \mu_X}{\sigma_X})$$

Where $\phi(x)$ is the standard normal cumulative distribution function (*cdf*).

Usually $X$ represents power variation measured in dB. Considering $X_{dB}$ with mean $\mu_{dB}$ and variance $\sigma_{dB}^2$, the corresponding lognormal RV $L = e^{\zeta X_{dB}} = 10^{\frac{X_{dB}}{10}}$ has the following pdf:

$$f_L(l, \mu_{dB}, \sigma_{dB}) = \begin{cases} \frac{1}{\sqrt{2\pi} l \sigma_{dB}} \exp(-\frac{1}{2\sigma_{dB}^2}[10\log(l) - \mu_{dB}]^2) & l > 0 \\ 0 & \text{otherwise} \end{cases} \quad (2)$$

Where $\mu_{dB} = \frac{\mu_X}{\xi}$ and $\xi = \frac{\ln(10)}{10}$

$\sigma_{dB} = \frac{\sigma_X}{\xi}$

The first two central moments of $L$ may be written as:

$$m = e^\mu e^{\sigma^2/2}$$
$$D^2 = e^{2\mu} e^{\sigma^2} (e^{\sigma^2} - 1) \qquad (3)$$

Correlated Lognormals sum distribution corresponds to the sum of dependent lognormal RVs, i.e.

$$\Lambda = \sum_{i=1}^{N} L_n = \sum_{i=1}^{N} e^{X_n} \qquad (4)$$

We define $\vec{L} = (L_1, L_2 ... L_N)$ as a strictly positive random vector such that the vector $\vec{X} = (X_1, X_2 ... X_N)$ with $X_j = \log(L_j)$, $1 \leq j < N$ has an n-dimensional normal distribution with mean vector $\mu = (\mu_1, \mu_2 ... \mu_N)$ and covariance matrix $M$ with $M(i,j) = Cov(X_i, X_j)$, $1 \leq i < N, 1 \leq j < N$. $\vec{L}$ is called an n-dimensional log-normal vector with parameters $\vec{\mu}$ and $M$.
$Cov(L_i, L_j)$ may be expressed as [9, eq. 44.35]:

$$Cov(L_i, L_j) = e^{\mu_i + \mu_j + \frac{1}{2}(\sigma_i^2 + \sigma_j^2)} (e^{M(i,j)} - 1) \qquad (5)$$

The first two central moments of $\Lambda$ may be written as:

$$\mathrm{m} = \sum_{i=1}^{N} m_i = \sum_{i=1}^{N} e^{\mu_i} e^{\sigma_i^2/2} \qquad (6)$$

$$D^2 = \sum_{i=1,j=1}^{N} Cov(L_i, L_j)$$
$$= \sum_{i=1,j=1}^{N} e^{\mu_i + \mu_j + \frac{1}{2}(\sigma_i^2 + \sigma_j^2)} (e^{M(i,j)} - 1) \qquad (7)$$

## 3. LOG SKEW NORMAL DISTRIBUTION

The standard skew normal distribution was firstly introduced in [10] and was independently proposed and systematically investigated by Azzalini [11]. The random variable $X$ is said to have a scalar $SN(\lambda, \varepsilon, \omega)$ distribution if its density is given by:

$$f_X(x; \lambda, \varepsilon, \omega) = \frac{2}{\omega} \varphi(\frac{x - \varepsilon}{\omega}) \phi(\lambda \frac{x - \varepsilon}{\omega}) \qquad (8)$$

Where 
$$\varphi(x) = \frac{e^{-x^2/2}}{\sqrt{2\pi}}, \qquad \phi(x) = \int_{-\infty}^{x} \varphi(\zeta) \, d\zeta$$

With $\lambda$ is the shape parameter which determines the skewness, $\varepsilon$ and $\omega$ represent the usual location and scale parameters and $\varphi$, $\phi$ denote, respectively, the *pdf* and the *cdf* of a standard Gaussian RV.
The CDF of the skew normal distribution can be easily derived as:

$$F_X(x; \lambda, \varepsilon, \omega) = \phi(\frac{x - \varepsilon}{\omega}) - 2 \mathrm{T}(\frac{x - \varepsilon}{\omega}, \lambda) \qquad (9)$$

Where function $\mathrm{T}(x, \lambda)$ is Owen's *T* function expressed as:

$$T(x,\lambda) = \frac{1}{2\pi} \int_0^\lambda \frac{\exp\left\{-\frac{1}{2}x^2(1+t^2)\right\}}{(1+t^2)} dt \qquad (10)$$

A fast and accurate calculation of Owen's T function is provided in [12].

Similar to the relation between normal and lognormal distributions, given a skew normal RV $X$ then $L = e^{\zeta X_{dB}} = 10^{\frac{X_{dB}}{10}}$ is a log skew normal distribution. The cdf and pdf of $L$ can be easily derived as:

$$f_L(l;\lambda,\varepsilon_{dB},\omega_{db}) = \begin{cases} \frac{2}{\xi\omega_{db}l}\varphi(\frac{10\log(l)-\varepsilon_{dB}}{\omega_{db}})\phi(\lambda\frac{10\log(l)-\varepsilon_{dB}}{\omega_{db}}) & l>0 \\ 0 \quad \text{otherwise} \end{cases} \qquad (11)$$

$$F_L(l;\lambda,\varepsilon_{dB},\omega_{db}) = \begin{cases} \phi(\frac{10\log(l)-\varepsilon_{dB}}{\omega_{db}}) - 2T(\frac{10\log(l)-\varepsilon_{dB}}{\omega_{db}},\lambda) & l>0 \\ 0 \quad \text{otherwise} \end{cases} \qquad (12)$$

Where $\varepsilon_{dB} = \frac{\varepsilon}{\xi}$ and $\xi = \frac{\ln(10)}{10}$

$\omega_{dB} = \frac{\omega}{\xi}$

The Moment Generating Function (*MGF*) of the skew normal distribution may be written as [11]:

$$M_X(t) = E\left[e^{tX}\right]$$
$$= 2e^{t^2/2}\phi(\beta t), \qquad \beta = \frac{\lambda}{\sqrt{1+\lambda^2}} \qquad (13)$$

Thus the first two central moments of $L$ are:

$$\zeta = 2e^\varepsilon e^{\omega^2/2}\phi(\beta\omega) \qquad (14)$$

$$\varpi^2 = 2e^{2\varepsilon}e^{\omega^2}(e^{\omega^2}\phi(2\beta\omega) - 2\phi^2(\beta\omega))$$

## 4. VALIDITY OF LOGNORMAL ASSUMPTION FOR THE SUM OF LOGNORMAL RVS AT DISTRIBUTION TAILS

Several approximation methods for the sum of correlated lognormal RVs is based on the fact that this sum can be approximated, at least as a first order, by another lognormal distribution. On the one hand, Szyszkowicz and Yanikomeroglu [14] have published a limit theorem that states that the distribution of a sum of identically distributed equally and positively correlated lognormal RVs converges, *in distribution*, to a lognormal distribution as N becomes large. This limit theorem is extended in [15] to the sum of correlated lognormal RVs having a particular correlation structure.

On the other hand, some recent results [16, Theorem 1. and 3.] show that the sum of lognormal RVs exhibits a different behaviour at the upper and lower tails even in the case of identically distributed lognormal RVs. Although the lognormal distribution have a symmetric behaviours in both tails, this is not in contradiction with results proven in [14-15] since convergence is proved *in distribution*, i.e., convergence at every point *x* not in the *limit behaviour*.

This explain why some lognormal based methods provide a good accuracy only in the lower tail (e.g. Schwartz and Yeh), where some other methods provide an acceptable accuracy in the upper tail (e.g. Fenton–Wilkinson). This asymmetry of the behaviours of the sum of lognormal RVs at the lower and upper tail, motivates us to use the Log Skew Normal distribution as it represents the asymmetric version of lognormal distribution. So, we expect that LSN approximation provide the needed accuracy over the whole region including both tails of the sum of correlated lognormal RVs distribution.

## 5. LOG SKEW NORMAL PARAMETERS DERIVATION

### 5.1. Tails properties of sum of Correlated lognormal RVs

Let $\vec{L}$ be an N-dimensional log-normal vector with parameters $\mu$ and $M$. Let $B = M^{-1}$ the inverse of the covariance matrix.

To study tails behavior of sum of correlated lognormal RVs, it is convenient to work on lognormal probability scale [13], i.e., under the transformation G:

$$G : F(x) \mapsto \widetilde{F}(x) = F(\phi^{-1}(F(e^x))) \quad (15)$$

We note that under this transformation, the lognormal distribution is mapped onto a linear equation.

We define $B_i$ as row sum of $B$:

$$B_i = \sum_{k=1}^{N} B(i,k), \quad 1 \leq i \leq N \quad (16)$$

We let:

$$\tilde{N} \triangleq Card\{i = 1,..N; \ B_i \neq 0\}$$
$$\tilde{I} \triangleq \{i = 1,..N; \ B_i \neq 0\} \triangleq \{\tilde{k}(1), \tilde{k}(1)...\tilde{k}(\tilde{N})\}$$

We define $\tilde{\mu}$, $\tilde{M}$, $\tilde{B}$ and $\tilde{B}_i$ such that:

$$\tilde{\mu}(i) = \mu(\tilde{k}(i))$$
$$\tilde{M}(i,j) = M(\tilde{k}(i), \tilde{k}(j))$$
$$\tilde{B} = \tilde{M}^{-1}$$
$$\tilde{B}_i = \sum_{k=1}^{\tilde{N}} \tilde{B}(i,k)$$

Since variables $L_i$ are exchangeable, we can assume for the covariance matrix B, with no loss of generality, that $\tilde{I} = \{1,2,3...\tilde{N}\}$ with $\tilde{N} \leq N$.

Let $w \in \Re^{\tilde{N}}$ such that:

$$w = \frac{\tilde{B}^{-1}1}{1^{\perp}\tilde{B}^{-1}1} \quad (17)$$

So that, we may write:

$$w_i = \frac{\tilde{A}_i}{\sum_{j=1}^{\tilde{N}} \tilde{A}_j} \quad j = 1,...\tilde{N} \quad (18)$$

Now, we set $\tilde{w} \in \Re^N$ as

$$\tilde{w}_i = \begin{cases} w_i & \text{if } i \leq \tilde{N} \\ 0 & \text{Otherwise} \end{cases} \quad (19)$$

Assuming that for every $i \in \{1,2,3...N\} \setminus \tilde{I}$

$$(e^i - \tilde{w})^\perp B\tilde{w} \neq 0 \quad (20)$$

Where $e^i \in \Re^N$ satisfies $e^i_j = 1$ if $i = j$ and $e^i_j = 0$ otherwise.

In [16, Theorem 3.], Gulisashvili and Tankov proved that the slope of the right tail of the SLN *cdf* on lognormal probability scale is equal to $1/\underset{i}{\text{Max}}\{\tilde{B}(i,i)\}$ when assumption (20) is valid.

$$\lim_{x \to +\infty} \frac{\delta}{\delta x} \widetilde{F}_{SCLN}(x) = \frac{1}{\underset{i}{\text{Max}}\{\tilde{B}(i,i)\}} \quad (21)$$

Considering the left tail slope, they proved that the slope of the left tail of the SLN *cdf* on lognormal probability scale is equal to $\sqrt{\sum_{i=1}^{N} \tilde{B}_i}$ [16, Theorem 1.].

$$\lim_{x \to -\infty} \frac{\delta}{\delta x} \widetilde{F}_{SCLN}(x) = \sqrt{\sum_{i=1}^{N} \tilde{B}_i} \quad (22)$$

In general, we can assume that $B_i \neq 0$ for $1 \leq i \leq N$, so that $N = \tilde{N}$ and tails slope can be expressed as:

$$\lim_{x \to +\infty} \frac{\delta}{\delta x} \widetilde{F}_{SCLN}(x) = \frac{1}{\underset{i}{\max}\{B(i,i)\}} \quad (23)$$

$$\lim_{x \to -\infty} \frac{\delta}{\delta x} \widetilde{F}_{SCLN}(x) = \sqrt{\sum_{i=1}^{N} B_i} \quad (24)$$

### 5.2. Tail properties of Skew Log Normal

In [17], it has been showed that the rate of decay of the right tail probability of a skew normal distribution is equal to that of a normal variate, while the left tail probability decays to zero faster. This result has been confirmed in [18]. Based on that, it is easy to show that the rate of decay of the right tail probability of a log skew normal distribution is equal to that of a lognormal variate. Under the transformation G, skew lognormal distribution has a linear asymptote in the upper limit with slope

$$\lim_{x \to +\infty} \frac{\delta}{\delta x} \widetilde{F}_{LSN}(x) = \frac{1}{w} \quad (25)$$

In the lower limit, it has no linear asymptote, but does have a limiting slope

$$\lim_{x \to -\infty} \frac{\delta}{\delta x} \widetilde{F}_{LSN}(x) = \frac{\sqrt{1+\lambda^2}}{w} \quad (26)$$

These results are proved in [8, Appendix A]. Therefore, it will be possible to match the tail slopes of the LSN with those of the sum of correlated lognormal RVs distribution in order to find LSN optimal parameters.

### 5.3. Moments and lower tail slope matching

In order to derive log skew normal optimal parameters, we proceed by matching the two central moments of both distributions. Furthermore, use we lower slope tail match. By simulation, we point out that upper slope tail match is valid only for the sum of high number of lognormals RVs. However we still need it to find an optimal starting guess solution to the generated nonlinear equation. Thus we define $\lambda_{opt}$ as solution the following nonlinear equation:

$$\frac{\sum_{i=1}^{N} e^{2\mu_i} e^{\sigma_i^2}(e^{\sigma_i^2}-1)}{(\sum_{i=1}^{N} e^{\mu_i} e^{\sigma_i^2/2})^2} = e^{\sqrt{\frac{1+\lambda^2}{\sum_{i=1}^{N}\tilde{B}_i}}} \frac{\phi(2\frac{\lambda}{\sqrt{\sum_{i=1}^{N}\tilde{B}_i}})\sqrt{\sum_{i=1}^{N}\tilde{B}_i}}{2\phi^2(\frac{\lambda}{\sqrt{\sum_{i=1}^{N}\tilde{B}_i}})} - 1 \quad (27)$$

Such nonlinear equation can be solved using different mathematical utility (e.g. fsolve in matlab). Using upper slope tail match we derive a starting solution guess $\lambda_0$ to (23) in order to converge rapidly (only few iterations are needed):

$$\lambda_0 = \sqrt{\left[\text{Max}_i\{\tilde{B}(i,i)\}^2 \sum_{i=1}^{N} \tilde{B}_i)\right] - 1} \quad (28)$$

Optimal location and scale parameters $\varepsilon_{opt}$, $\omega_{opt}$ are obtained according to $\lambda_{opt}$ as.

$$\begin{cases} \omega_{opt} = \sqrt{\frac{1+\lambda_{opt}^2}{\sum_{i=1}^{N}\bar{B}_i}} \\ \varepsilon_{opt} = \ln(\sum_{i=1}^{N} e^{\mu_i} e^{\sigma_i^2/2}) - \frac{\omega_{opt}^2}{2} - \ln(\phi(\frac{\lambda_{opt}}{\sqrt{\sum_{i=1}^{N}\bar{B}_i}})) \end{cases} \quad (29)$$

### 6. SIMULATION RESULTS

In this section, we propose to validate our approximation method and compare it with other widely used approximation methods. The comparison of the Complementary Cumulative Distribution Function (*CDF*) of the sum of $N$ ($N=2, 8, 20$) correlated lognormal RVs $P[\Lambda > \lambda]$ of Monte Carlo simulations with LSN approximation and lognormal based methods for low value of standard deviation $\sigma = 3\,dB$ with $\mu = 0\,dB$ and $\rho = 0.7$ are shown in Fig.1. Although these methods are known by its accuracy in the lower range of standard deviation, it obvious that LSN approximation outperforms them. We note that fluctuation at the tail of sum of lognormal RVs distribution is due to Monte Carlo simulation, since we consider $10^7$ samples at every turn. We can see that LSN approximation results are identical to Monte Carlo simulation results.

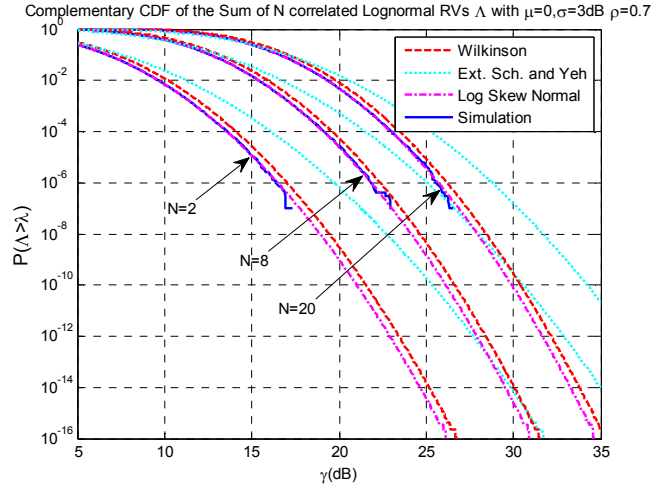

Figure 1. Complementary CDF of the sum of N correlated lognormal RVs with $\mu = 0\,dB$, $\sigma = 3\,dB$ and $\rho = 0.7$

Fig.2 and Fig.3 show the complementary *CDF* of the sum of *N* correlated lognormal RVs for higher values of standard deviation $\sigma = 6, 9\,dB$ and two values of correlation coefficients $\rho = 0.9, 0.3$. We consider the Log Shifted Gamma approximation for comparison purposes. We consider the Log Shifted Gamma approximation for comparison purposes. We can see that LSN approximation highly outperforms other methods especially at the *CDF* right tail ($1 - CDF < 10^{-2}$). Furthermore, LSN approximation give exact Monte Carlo simulation results even for low range of the complementary CDF of the sum of correlated lognormal RVs ($1 - CDF < 10^{-6}$).

To further verify the accuracy of our method in the left tail as well as the right tail of *CDF* of the sum of correlated lognormal RVs, Fig. 4 and Fig. 5 show respectively the *CDF* $P[\Lambda < \lambda]$ of the sum of 6 correlated lognormal RVs with $\mu = 0\,dB$ $\rho = 0.7$ and the complementary *CDF* of the sum of 20 correlated lognormal RVs with $\mu=0dB$ $\rho=0.3$ for different standard deviation values. It is obvious that LSN approximation provide exact Monte Carlo simulation results at the left tail as well as the right tail. We notice that accuracy does not depend on standard deviation values as LSN performs well for the highest values of standard deviation of the shadowing.

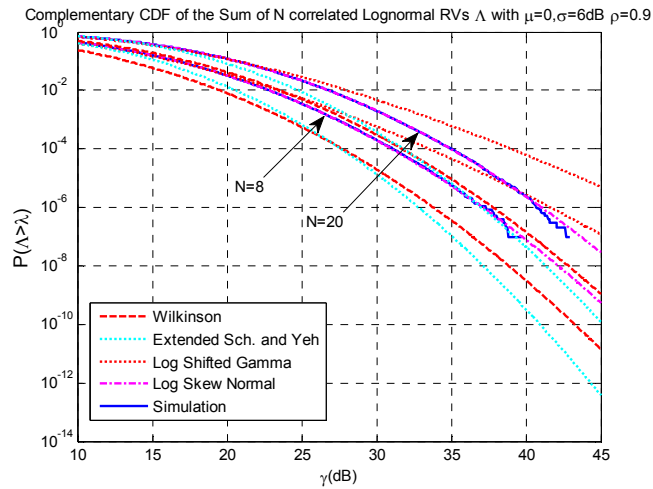

Figure 2. Complementary CDF of the sum of N correlated lognormal RVs with $\mu = 0\,dB$, $\sigma = 6\,dB$ and $\rho = 0.9$

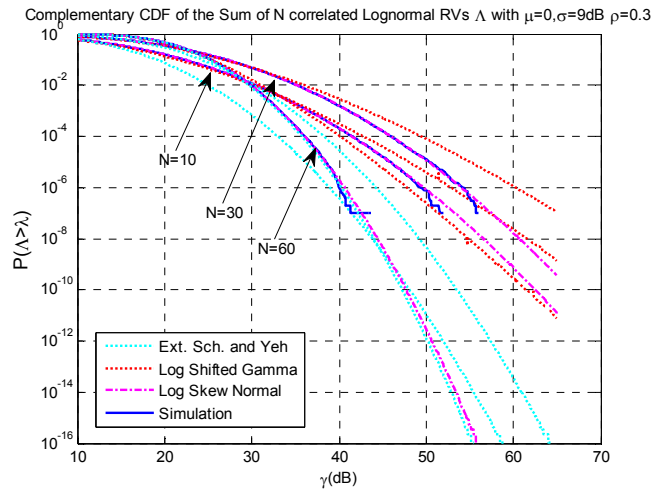

Figure 3. Complementary CDF of the sum of N correlated lognormal RVs with $\mu = 0\,dB$ $\sigma = 9\,dB$ $\rho = 0.3$

To point out the effect of correlation coefficient value on the accuracy of the proposed approximation, we consider the *CDF* of the sum of 12 correlated lognormal RVs for different combinations of standard deviation and correlation coefficient values with $\mu=0dB$ (Fig. 6). One can see that LSN approximation efficiency does not depend on correlation coefficient or standard deviation values. So that, LSN approximation provides same results as Monte Carlo simulations for all cases.

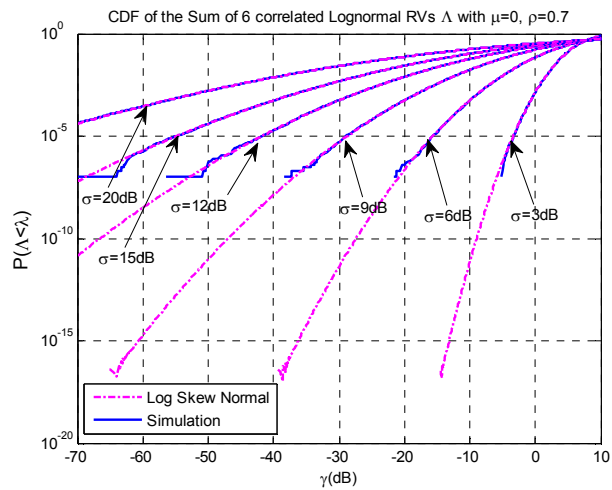

Figure 4. CDF of the sum of 6 correlated lognormal RVs with $\mu = 0\,dB$, $\rho = 0.7$

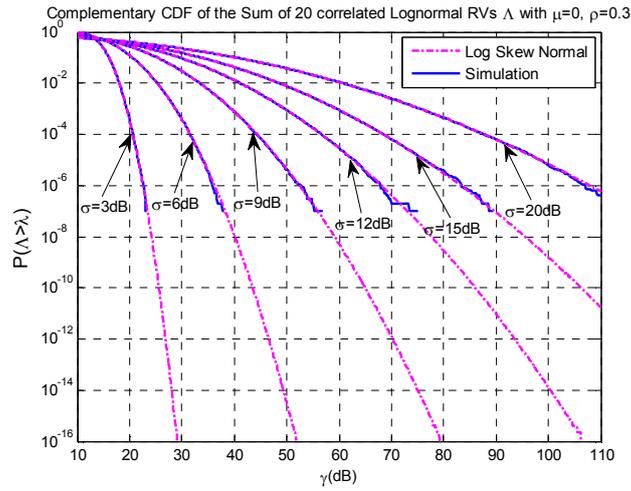

Figure 5. Complementary CDF of the sum of 20 correlated lognormal RVs with $\mu = 0\,dB$, $\rho = 0.3$

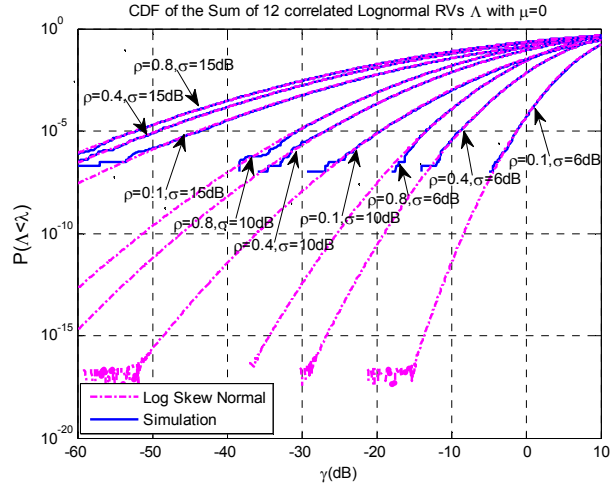

Figure 6. CDF of the sum of 12 correlated lognormal RVs with different correlation coefficients $\rho$, $\mu = 0\,dB$

## 7. CONCLUSIONS

In this paper, we proposed to use the Log Skew Normal distribution in order to approximate the sum of correlated lognormal RV distribution. Our fitting method uses moment and tails slope matching technique to derive LSN distribution parameters. LSN provides identical results to Monte Carlo simulations results and then outperforms other methods for all cases. Such approximation can be effectively used for accurate and fast computation of interference and outage probabilities in cellular networks.

# REFERENCES

[1] A. Safak, "Statistical analysis of the power sum of multiple correlated log-normal components", IEEE Trans. Veh. Tech., vol. 42, pp. 58–61, Feb. 1993.

[2] S.Schwartz and Y.S. Yeh, "On the distribution function and moments of power sums with log-normal components", Bell System Tech. J., Vol. 61, pp. 1441–1462, Sept. 1982.

[3] M. Pratesi , F. Santucci , F. Graziosi and M. Ruggieri "Outage analysis in mobile radio systems with generically correlated lognormal interferers", IEEE Trans. Commun., vol. 48, no. 3, pp.381 -385 2000.

[4] A. Safak and M. Safak "Moments of the sum of correlated log-normal random variables", Proc. IEEE 44th Vehicular Technology Conf., vol. 1, pp.140 -144 1994.

[5] C. L. Joshua Lam,Tho Le-Ngoc " Outage Probability with Correlated Lognormal Interferers using Log Shifted Gamma Approximation" , Wireless Personal Communications, Volume 41, Issue 2, pp 179-192, April 2007.

[6] Z. Wu, X. Li, R. Husnay, V. Chakravarthy, B. Wang, and Z. Wu. A novel highly accurate log skew normal approximation method to lognormal sum distributions. In Proc. of IEEE WCNC 2009.

[7] ] X. Li, Z. Wu, V. D. Chakravarthy, and Z. Wu, "A low complexity approximation to lognormal sum distributions via transformed log skew normal distribution," IEEE Transactions on Vehicular Technology, vol. 60, pp. 4040-4045, Oct. 2011.

[8] M. Benhcine, R. Bouallegue, "Fitting the Log Skew Normal to the Sum of Independent Lognormals Distribution", accepted in the sixth International Conference on Networks and Communications (NetCom-2014) .

[9] Samuel Kotz, N. Balakrishnan, Norman L. Johnson "Continuous Multivariate Distributions", Volume 1, Second Edition, John Wiley & Sons, avril 2004.

[10] O'Hagan A. and Leonard TBayes estimation subject to uncertainty about parameter constraints, Biometrika, 63, 201–202, 1976.

[11] Azzalini A, A class of distributions which includes the normal ones, Scand. J. Statist., 12, 171–178, 1985.

[12] M. Patefield, "Fast and accurate calculation of Owen's t function," J. Statist. Softw., vol. 5, no. 5, pp. 1–25, 2000.

[13] N. C. Beaulieu, Q. Xie, "Minimax Approximation to Lognormal Sum Distributions", IEEE VTC Spring, Vol. 2, pp. 1061-1065, April 2003.

[14] S. S. Szyszkowicz and H. Yanikomeroglu "Limit theorem on the sum of identically distributed equally and positively correlated joint lognormals", IEEE Trans. Commun., vol. 57, no. 12, pp.3538 -3542 2009

[15] N. C. Beaulieu, "An extended limit theorem for correlated lognormal sums," IEEE Trans. Commun., vol. 60, no. 1, pp. 23-26, Jan. 2012

[16] Archil Gulisashvili, Peter Tankov, "Tail behavior of sums and differences of log-normal random variables", ARXIV  09/2013.

[17] Antonella Capitanio, "On the approximation of the tail probability of the scalar skew-normal distribution", in METRON (2010).

[18] W. Hürlimann, "Tail Approximation of the Skew-Normal by the Skew-Normal Laplace: Application to Owen's T Function  and the Bivariate Normal Distribution", Journal of Statistical and Econometric Methods, vol. 2, no.1, 2013, 1-12, Scienpress Ltd, 2013.

## Authors

**Marwane Ben Hcine** was born in Kébili, Tunisia, on January 02, 1985. He graduated in Telecommunications Engineering, from The Tunisian Polytechnic School (TPS), July 2008. In June 2010, he received the master's degree of research in communication systems of the Higher School of Communication of Tunisia (Sup'Com). Currently he is a Ph.D. student at the Higher School of Communication of Tunisia. His research interests are network design and dimensioning for LTE and beyond Technologies.

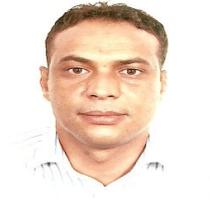

**Pr. Ridha BOUALLEGUE** was born in Tunis, Tunisia. He received the M.S degree in Telecommunications in 1990, the Ph.D. degree in telecommunications in 1994, and the HDR degree in Telecommunications in 2003, all from National School of engineering of Tunis (ENIT), Tunisia. Director and founder of National Engineering School of Sousse in 2005. Director of the School of Technology and Computer Science in 2010. Currently, Prof. Ridha Bouallegue is the director of Innovation of COMmunicant and COoperative Mobiles Laboratory, INNOV'COM Sup'COM, Higher School of Communication. His current research interests include mobile and cooperative communications, Access technique, intelligent signal processing, CDMA, MIMO, OFDM and UWB systems.

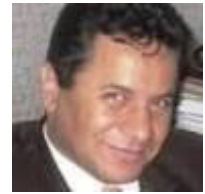